\documentclass[letterpaper]{jpconf}
\usepackage{graphicx}
\usepackage{bm}
\usepackage{eufrak}
\usepackage{slashed}
\usepackage{amsmath}
\usepackage{epsfig}
\newcommand{\exclude}[1]{}
\begin{document}
\title{The Axial Anomaly and Large Pulsar Kicks\footnote{presented at the 25th Lake Louise Winter Institute under the title ``Kicking Pulsars Hard''}  }

\author{James Charbonneau}

\address{University of British Columbia, \em{james@phas.ubc.ca}}

\begin{abstract} 
Topological vector currents have gained interest recently with their possible verification at RHIC through the Charge Separation Effect and the Chiral Magnetic Effect. Much work has been done in understanding the role of topological vector currents in astrophysics, specifically in the interiors of neutron stars and quark stars. We will discuss a recent aspect of this work regarding pulsar kicks. A significant percentage of the pulsar population is known to have velocities above 1000 km/s, but a suitable explanation for these velocities does not exist. We will detail how topological currents may be responsible for these large kicks and discuss why the mechanism is successful where others fail. 

\end{abstract}

\section{Introduction}
A recent topic of much interest has been the $\mathcal{P}$ and $\mathcal{CP}$-odd effects that arise from the axial anomaly. The most popular of these has been the Chiral Magnetic Effect \cite{Fukushima:2008xe}, but this is part a body of work investigating this phenomenon that starts with topological currents in condensed matter systems \cite{Alekseev:1998ds}, and includes the study of anomalous axion interactions in QCD \cite{Metlitski:2005pr}, the Charge Separation Effect \cite{Kharzeev:2007tn}, and the high density analogue of the Chiral Magnetic Effect in dense stars \cite{Charbonneau:2007db,Charbonneau:2009ax}. The Chiral Magnetic effect is particularly exciting because it rests on the edge of observational science. The current may be responsible for the parity violating effects seen in the STAR collaboration at RHIC \cite{Voloshin:2009hr}. Here we will discuss how the existence of these currents in dense stars may be responsible for generating the large proper motion seen in some pulsars \cite{Charbonneau:2009hq}.

The goal of the paper \cite{Charbonneau:2009hq} was to elaborate on a kick mechanism first discussed by \cite{Charbonneau:2009ax} that may explain pulsar velocities greater than $1000$~km$\,$s$^{-1}$. There have been a number of studies that have compiled and modelled the velocities of pulsars. Although they disagree on whether the distribution is indeed bimodal, they agree that a significant number of pulsars are travelling faster than can be attributed to neutrino kicks. The analysis of \cite{Arzoumanian:2001dv} favours a bimodal velocity distribution with peaks at 90~km$\,$s$^{-1}$ and 500~km$\,$s$^{-1}$ with $15\%$ of pulsars travelling at speeds greater than 1000 km$\,$s$^{-1}$. Alternatively \cite{Hobbs:2005yx} and \cite{0004-637X-643-1-332} both predict a single peaked distribution with an average velocity of $\sim 400$~km$\,$s$^{-1}$, but point out that the faster pulsars B2011+38 and B2224+64 have speeds of $\sim 1600$~km$\,$s$^{-1}$. Large velocities are unambiguously confirmed with the model independent measurement of pulsar B1508+55 moving at $1083^{+103}_{-90}$~km$\,$s$^{-1}$ \cite{Chatterjee:2005mj}.

Currently no mechanism exists that can reliably kick the star hard enough to reach these velocities.  Asymmetric explosions can only reach 200~km$\,$s$^{-1}$ \cite{Fryer:2003tc}, and asymmetric neutrino emission is plagued by the problem that at temperatures high enough to produce the kick the neutrinos are trapped inside the star \cite{Sagert:2007ug}. Alterations of the neutrino model that take into account only a thin shell of neutrinos require large temperatures and huge surface magnetic fields.

\section{Generating Large Kicks}
We will provide a sketch of how the kick is generated and direct those interested in the details to read \cite{Charbonneau:2009hq}. The kick mechanism we will discuss relies on the existence of topological vector currents of the form described by \cite{Charbonneau:2009ax}, which some readers may recognize as the same current responsible for the Chiral Magnetic Effect \cite{Fukushima:2008xe} in QCD, 
\begin{eqnarray}  
\label{j}
\langle j \rangle = (n_L - n_R)\frac{e \Phi}{2\pi^2},
\end{eqnarray}  
where $n_R$ and $n_L$ are the one dimensional number densities of the right and left-handed electrons, and $\Phi$ is the magnetic flux.   There are three requirements for topological vector currents to be present: an imbalance in left and right-handed particles $\mu_L\neq \mu_R$, degenerate matter $\mu\gg T$, and the presence of a background magnetic field $B\neq 0$.  All of these are present in neutron and quark stars.  The weak interaction, by which the star attains equilibrium, violates parity; particles created in this environment are primarily left-handed.  The interior of the star is very dense, $\mu_e \sim 10$~MeV, and cold, $T\sim0.1$~MeV, such that the degeneracy condition $\mu\gg T$ is met, and neutron stars are known to have huge surface magnetic fields, $B_\textrm{s}\sim10^{12}$~G.

If the electrons carried by the current can transfer their momentum into space---either by being ejected or by radiating photons---the current could push the star like a rocket.  In typical neutron stars this is unlikely because the envelope (the region where $\mu\sim T$) is thought to be about 100~m thick. Once the current reaches this thick crust, it will likely be reabsorbed into the bulk of the star. But if the crust is very thin, or nonexistent, the electrons may leave the system or emit photons that will carry their momentum to space. The electrosphere for bare quark stars is thought to be about 1000~fm. With this in mind we conjecture that stars with very large kicks, $v\gg 200$~km$\,$s$^{-1}$, are quark stars and that slow moving stars, $v \leq 200$~km$\,$s$^{-1}$, are kicked by some other means, such as asymmetric explosions or neutrino emission, and are typical neutron stars.  Confirmation of this would provide an elegant way to discriminate between neutron stars and quark stars.

The total number current for electrons reaching the surface of the star is calculated in  \cite{Charbonneau:2009hq} and is given by,
\begin{eqnarray}
\label{eq:current_quark}
\langle j \rangle =  2.6 \times 10^{26}\left( \frac{10 \textrm{ MeV}}{k_e} \right) \left( \frac{B_\textrm{core}}{10\, B_\textrm{c}} \right) \left(\frac{T_\textrm{core}}{10^9 \textrm{ K}}\right)^{5} \left(\frac{n_\textrm{b}}{n_0}\right)^{1/3} \textrm{MeV}\,, 
\end{eqnarray}
where $B_\textrm{c} = 4.4\times10^{13}$ G is the critical magnetic field, $T_\textrm{core}$ is the core temperature of the star, and $n_0$ is nuclear density. The typical density for quark matter is $n_\textrm{b} \sim 10\, n_0$ but could easily be higher. Though many pulsars have a surface field of around $10^{12}$ G,  the field in the bulk of the star is likely much stronger based on virial theorem arguments in \cite{Lai:1991},which yield possible core fields of $B_\textrm{max} \sim10^{18}$ G. This is an extremely large field and is unlikely as it is a strict upper bound.  Based on this we choose a value of the core magnetic field to be $B_\textrm{core}=10\,B_\textrm{c}$. 

\begin{figure} 
\begin{center}
 \includegraphics[width=7cm]{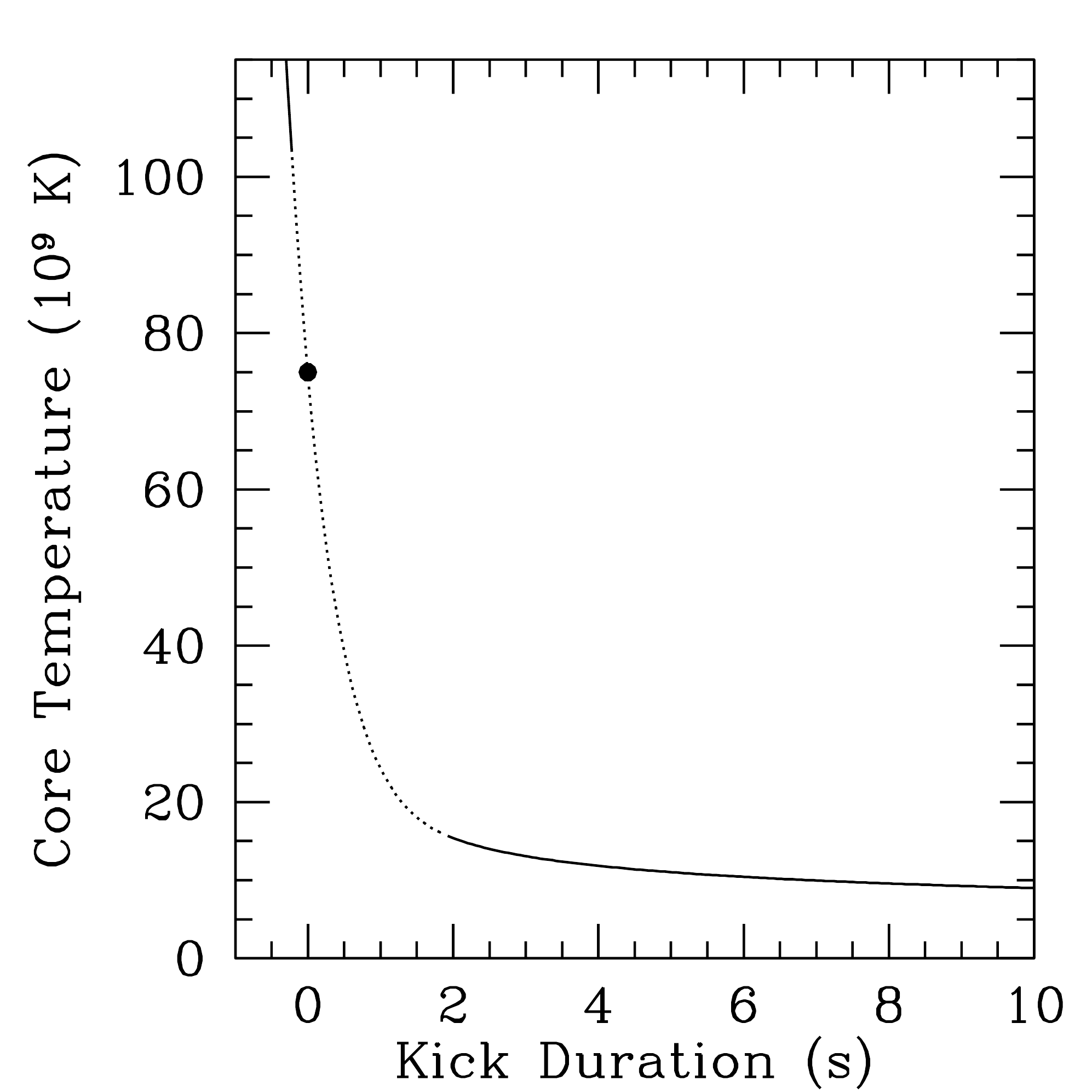} \hspace{1.5cm} \includegraphics[width=7cm]{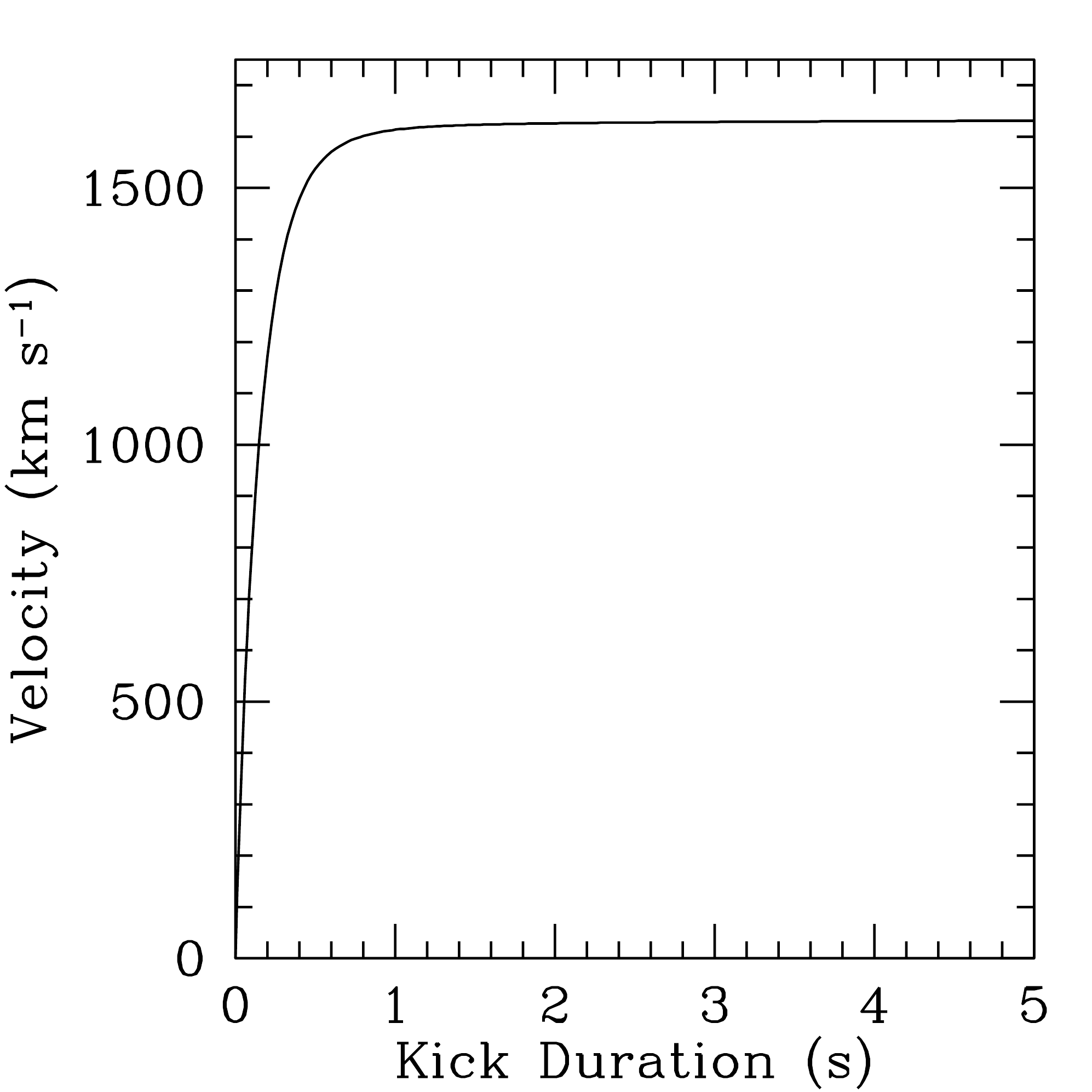}
  \caption{{\em a)} The dashed line indicates the translucent part of the cooling curve, modelled by exponential decay. The curve before the patch is taken from \cite{Haensel:1991um} and the curve after the patch is from  \cite{Page:2002bj}. The black dot marks the start of the kick at $t=0$. {\em b)} Time evolution of the kick for an internal magnetic field $B=10\, B_\textrm{c}$.}
  \label{figure}
\end{center}  
\end{figure}

The current, and thus the kick, is very sensitive to the cooling of the star. Unfortunately, kicks are likely to occur right after the birth of the star during the most poorly understood stage of cooling. The initial cooling of the star is described in \cite{Haensel:1991um}, which focuses on neutrino diffusion through the star and thermal cooling. The star then cools until the neutrinos can escape the quark star and the cooling moves into a purely radiative regime as discussed in \cite{Page:2002bj}.  The part of the cooling curve between these two well defined mechanisms constitutes the translucent regime, which we model with and exponential decay as shown in Figure~\ref{figure}a.

The degeneracy of the electrons is responsible for powering the kick. Each electron carries a momentum equal to its Fermi momentum, which is quite large due to the extreme degeneracy in the star. As seen in Figure \ref{figure}b, the star quickly reaches a speed of $v_{\textrm{max}} \sim 1600 \textrm{ km$\,$s$^{-1}$}$, which is big enough to account for the large kicks seen in many pulsars. As plotted, the entire kick seems to happen very quickly, but the current keeps running throughout the star's life. With a large internal magnetic field the mechanism can account for kicks seen in young pulsars.  But because the kick is constantly running, pulsars with smaller internal magnetic fields will eventually attain very large speeds very late in life.

\section{The Difference between Topological Kicks and Neutrino Kicks}
Neutrino kicks and topological kicks seem very similar on the surface. The electrons and neutrinos that contribute towards their kicks are created at the same rate $w$, have nearly the same degree of helicity, and have the same occupation of the lowest Landau level $n_\textrm{L}$. This means the flux of particles contributing to both the electron kick and the neutrino kick is about the same $\sim n_\textrm{L}\,w$. The difference between the two mechanisms comes from the momentum that the relevant particle carries. The neutrinos are created thermally and the typical momentum of a neutrino is equal to the temperature of the star $T$. The momentum of the electrons comes from the large chemical potential, $\mu_e\sim 10$ MeV. The momentum transfer per unit time for neutrinos is $F_\nu \sim T n_\textrm{L} w$ and for electrons is $F_e \sim \mu_e n_\textrm{L} w$. When the kick starts the star has a temperature of only $T\sim 1$ MeV. The electron kick is stronger than the neutrino kick by a factor of
\begin{eqnarray}
\frac{F_e}{F_\nu} \sim  \frac{\mu_e } {T}\,.
 \end{eqnarray}
Initially, when the star is very hot, the electron kick is an order of magnitude stronger than the neutrino kick. Furthermore, as the star cools the neutrino kick gets even weaker, while the electrons continue to have a momentum dictated by their chemical potential. This is how electrons generate larger kicks than neutrinos in a similar environment.

\break
\section{The Affect of the Current on the Cooling of the Star}
At the beginning of the star's life the energy from the kick does not contribute to the cooling of the star, but later in life the current could over take neutrino cooling as the dominant mechanism. 

This is because only a small fraction of the electrons created in the star actually escape, whereas all the neutrinos created in the star escape. The electrons only propagate because the asymmetry in the lowest Landau level and detailed balance allows the helicity states to propagate out of the star.  Those electrons that do not contribute toward the kick are trapped inside the star. Only those helicity states that reach the surface contribute to the cooling of the star. The neutrinos cool the star with a luminosity $L_\nu \sim T w$ where the electrons cool the star with an energy current (luminosity) of $L_e \sim \mu_e n_\textrm{L} w$. The ratio of electron cooling to neutrino cooling is
\begin{eqnarray}
\frac{L_e}{L_\nu}\sim\frac{\mu_e n_L} {T} =  \frac{m_e^2}{\mu_e T} \frac{B}{B_c}\,.
 \end{eqnarray}
At first the electrons cool the star at about 1/100 the rate of neutrino cooling. As the star cools, eventually $L_e/L_\nu > 1$ and the more energy is lost due to the current than the neutrinos. This transition occurs at a temperature
\begin{eqnarray}
T_\textrm{t}\sim 10^{-2} \left(\frac{B}{B_{\textrm{c}}}\right)\textrm{ MeV} \sim10^8  \left(\frac{B}{B_{\textrm{c}}}\right) \textrm{ K}\,,
 \end{eqnarray} 
well after the kick has occurred. The current may be an additional cooling mechanism to consider in stars that have cooled below $10^8$ K. 

\ack
This work was supported by the National Science and Engineering Council of Canada.

\section*{References}
%\bibliography{mybib}

\providecommand{\newblock}{}

\end{document}